\documentclass[11pt]{article}

\usepackage[margin=1in]{geometry}
\usepackage[T1]{fontenc}
\usepackage[utf8]{inputenc}
\usepackage{lmodern}
\usepackage{microtype}
\usepackage{graphicx}
\usepackage{booktabs}
\usepackage{tabularx}
\usepackage{array}
\usepackage{enumitem}
\usepackage{xurl}
\usepackage{tikz}
\usepackage{float}
\usepackage{hyperref}
\usepackage[numbers,sort&compress]{natbib}

\usetikzlibrary{arrows.meta,positioning,calc,fit,backgrounds,shapes.geometric,decorations.pathmorphing}

\hypersetup{
  colorlinks=true,
  linkcolor=blue,
  citecolor=blue,
  urlcolor=blue
}

\newcolumntype{Y}{>{\raggedright\arraybackslash}X}

\definecolor{accent}{HTML}{2563EB}
\definecolor{accentsoft}{HTML}{EAF2FF}
\definecolor{accentsofttwo}{HTML}{F8FAFC}
\definecolor{accentdeep}{HTML}{1D4ED8}
\definecolor{slate}{HTML}{1E293B}
\definecolor{muted}{HTML}{64748B}
\definecolor{sand}{HTML}{FFF7ED}
\definecolor{warmline}{HTML}{FDBA74}
\definecolor{capsule}{HTML}{FEF3C7}
\definecolor{capsuleline}{HTML}{D97706}

\tikzset{
  accentarrow/.style={-{Latex[length=2.6mm,width=1.8mm]}, draw=accentdeep, line width=1.05pt},
  softarrow/.style={-{Latex[length=2.2mm,width=1.5mm]}, draw=black!38, line width=0.85pt},
  meshlink/.style={draw=accent!42, line width=0.95pt},
  legionlink/.style={draw=accent!55, line width=0.7pt, decorate, decoration={snake, amplitude=0.35mm, segment length=2.6mm}},
  moderncard/.style={draw=black!13, rounded corners=6pt, line width=0.8pt, fill=white, inner sep=9pt, align=left, font=\small\sffamily, text=slate},
  warmcard/.style={draw=warmline!65!black, rounded corners=10pt, line width=0.9pt, fill=sand, inner sep=8pt},
  bandcard/.style={draw=black!12, rounded corners=6pt, line width=0.8pt, fill=accentsofttwo, inner sep=8pt},
  stagecard/.style={draw=black!13, rounded corners=6pt, line width=0.8pt, fill=white, inner sep=8pt, align=left, font=\small\sffamily, text=slate, minimum height=2.25cm},
  stagebadge/.style={draw=none, circle, fill=accent, text=white, minimum size=18pt, inner sep=0pt, font=\scriptsize\sffamily\bfseries},
  figuretag/.style={draw=none, rounded corners=8pt, fill=accentdeep, text=white, inner xsep=6pt, inner ysep=2pt, font=\scriptsize\sffamily\bfseries},
  chip/.style={draw=none, rounded corners=8pt, fill=accentsoft, text=accentdeep, inner xsep=6pt, inner ysep=3pt, font=\scriptsize\sffamily\bfseries},
  mutedchip/.style={draw=none, rounded corners=8pt, fill=white, text=muted!85!black, inner xsep=6pt, inner ysep=3pt, font=\scriptsize\sffamily\bfseries},
  ghostpanel/.style={draw=black!10, rounded corners=6pt, line width=0.8pt, fill=black!1, inner sep=8pt},
  legacymini/.style={draw=black!14, rounded corners=4pt, line width=0.75pt, fill=white, inner sep=4pt, align=center, font=\scriptsize\sffamily, text=slate},
  meshpanel/.style={draw=black!12, rounded corners=6pt, line width=0.85pt, fill=accentsofttwo, inner sep=8pt},
  legionagent/.style={draw=black!16, rounded corners=4pt, line width=0.8pt, fill=white, inner sep=4pt, align=center, font=\scriptsize\sffamily, text=slate, minimum width=1.55cm},
  capsulebadge/.style={draw=capsuleline, rounded corners=2pt, line width=0.55pt, fill=capsule, inner xsep=2.5pt, inner ysep=1pt, font=\fontsize{5.6pt}{6.5pt}\sffamily, text=capsuleline!80!black},
  meshcore/.style={draw=warmline!70!black, ellipse, line width=1.0pt, fill=sand, inner sep=8pt, align=center, font=\small\sffamily\bfseries, text=slate}
}

\title{Hephaestus: Toward a Cybersecurity AI Scientist\thanks{\emph{Hephaestus}, the divine artisan of Homeric tradition, is the maker of both spear and shield. The name marks a research system that produces both offensive and defensive scientific artifacts under one design.}}
\author{Jiaqi Li, Yang Zhao, Wen Lu, Lvyang Zhang, and Lidong Zhai\\
Institute of Information Engineering, Chinese Academy of Sciences, Beijing, China\\
School of Cyber Security, University of Chinese Academy of Sciences, Beijing, China\\
\texttt{zhailidong@iie.ac.cn}}
\date{June 26, 2026}

\begin{document}
\maketitle

\begin{abstract}
Cyber offense is moving to machine speed; cyber research itself is not. Existing AI scientist systems make end-to-end research automation increasingly plausible, but they target relatively stable scientific domains. We argue that AI-native cybersecurity is a different kind of scientific object. Its recurring units of study are security events and interaction traces, not static assets; its model and tool substrate is non-stationary, not steady-state; and credible evaluation depends on digital twins, cyber ranges, and auditable evidence rather than on a single benchmark score. We call this object the Cybersecurity AI Scientist. A practical realization is a modular, role-specialized multi-agent research system that coordinates problem framing, threat modeling, tool generation, controlled experimentation, evaluation, governance, and scientific reporting, and that anchors its concrete objectives in a four-zeros frame spanning risk, trust, incident, and energy dimensions. As a representative agenda we focus on AI-native defense, where steady-state perimeters give way to resilient agent legions and the classical category of terminal security is itself being deconstructed into agent security. This paper defines the object, separates it from any single organizational realization, and offers an architecture and an agenda on which later systems, benchmarks, and empirical programs can be built.
\end{abstract}

\section{Introduction}

Two trajectories have begun to converge on cybersecurity. On one side, autonomous AI agents are already pushing into operational territory: language-model agents exploit real one-day vulnerabilities, and team-of-agent systems improve performance on harder zero-day scenarios \citep{one_day_vuln_2404_08144,zero_day_team_2406_01637}. On the other side, AI scientist systems are pushing research itself toward automation, with end-to-end pipelines that generate ideas, write code, run experiments, draft papers, and conduct simulated review \citep{ai_scientist_2408_06292,ai_scientist_v2_2504_08066}, and with domain scientist platforms in biology and biomedicine that organize multi-agent hypothesis generation, experiment planning, and analysis \citep{ai_co_scientist_nature_2026,robin_nature_2026,paperqa2_2409_13740,era_2509_06503}.

These two trajectories have so far progressed in parallel. AI scientist work has matured on relatively stable scientific objects, where the system under study does not adapt to the act of being studied. Cybersecurity agent work, in turn, has concentrated on task execution and benchmark performance, ranging from autonomous penetration testing and exploitation to evaluation of cyber-expert capability and the security of the agents themselves \citep{survey_agentic_cyber_2601_05293,digital_cyber_expert_2504_11783,co_redteam_2602_02164,harness_vuln_discovery_2604_20801,security_of_ai_agents_2406_08689}. Cybersecurity has therefore become a frontier arena for AI capability and control without yet being framed as a scientist-level research object in its own right. Recent institutional efforts, including Google's Big Sleep and DARPA's AI Cyber Challenge, make clear that autonomous cyber reasoning and AI-native defense are no longer hypothetical \citep{google_summer_security_2025,aixcc_program_2023,aixcc_results_2025}.

We argue that this gap is not bridged by a simple domain transfer. The cybersecurity research substrate is unusually unforgiving for generic AI scientist pipelines. Adversaries adapt to the very capabilities used to study them. Dual-use pressure is structural rather than peripheral. Model platforms, guardrails, and tool affordances drift on a timescale shorter than the research loop. And the validity of conclusions depends on environments, digital twins, and evidence chains that are themselves part of the method. What is needed is not a renamed agent pipeline but a cybersecurity-specific research object.

We call this object the \emph{Cybersecurity AI Scientist}. It denotes the kind of research capability that becomes necessary when scientific-problem shift and research-paradigm shift in cybersecurity are taken seriously at the same time. At the systems level, one practical realization is a modular, role-specialized multi-agent research system that coordinates problem framing, literature and threat analysis, tool generation, digital-twin experimentation, evaluation, governance-aware analysis, and scientific reporting through specialized agents, reusable capability packages, and bounded tools. We adopt and extend the end-to-end scientific pipeline of The AI Scientist line of work \citep{ai_scientist_2408_06292,ai_scientist_v2_2504_08066}: we keep the closed-loop ideation-execution-reporting structure, but we replace its assumption of a steady computational substrate with an adversarial, non-stationary, multi-model substrate, and we make governance and environment design first-class parts of the method.

\begin{figure}[t]
\centering
\includegraphics[width=\linewidth]{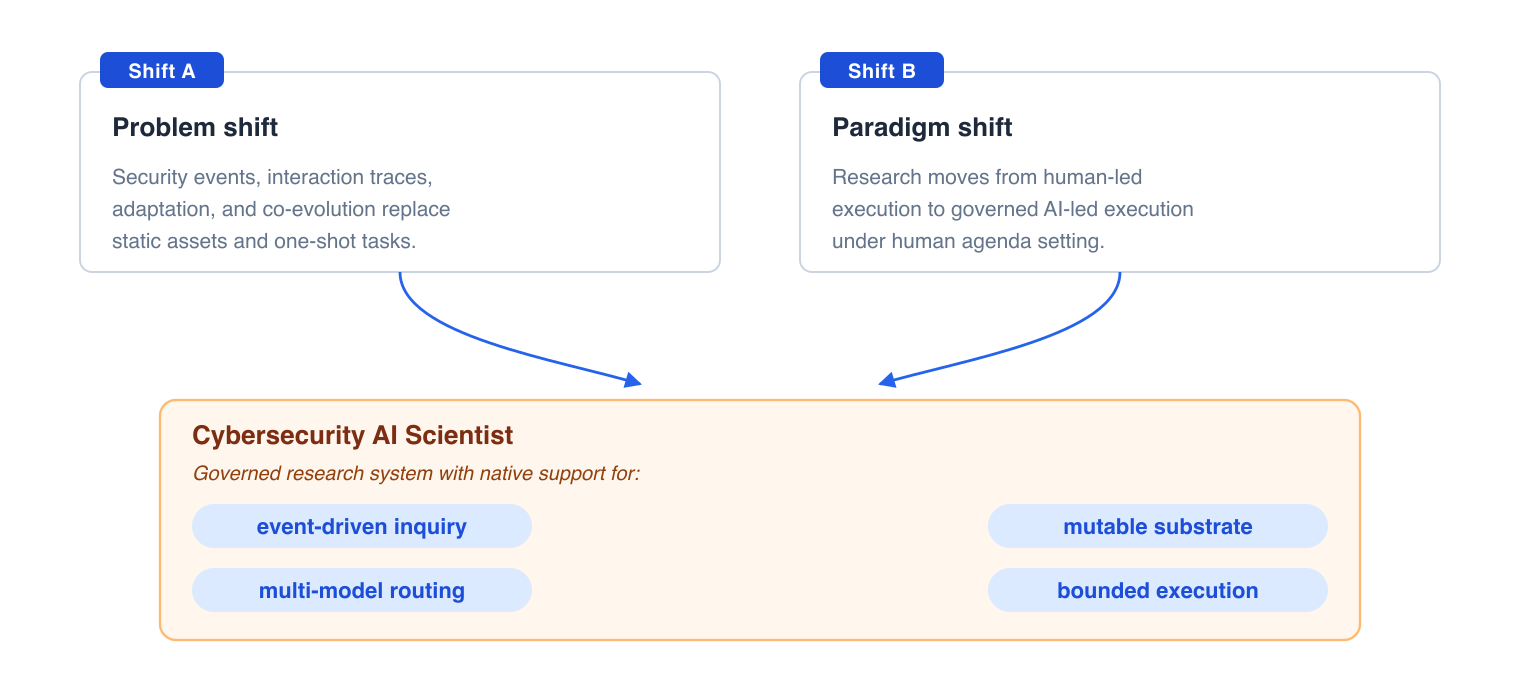}
\caption{Cybersecurity AI Scientist is motivated by two coupled shifts. The scientific object changes, and the organization of research changes with it.}
\label{fig:shift}
\end{figure}

\paragraph{Contributions.}
\begin{enumerate}[leftmargin=1.5em]
  \item We define the Cybersecurity AI Scientist as a distinct research object rather than a domain-specific relabeling of generic AI scientist systems.
  \item We characterize the coupled problem shift and research-paradigm shift induced by AI-native security conditions, and we anchor concrete objectives in a four-zeros frame spanning risk, trust, incident, and energy dimensions.
  \item We give a framework architecture and research pipeline, and present a modular multi-agent research system organized around specialized roles and reusable capability packages as one plausible organizational realization.
  \item We identify AI-native defense, including resilient agent legions and the deconstruction of terminal security into agent security, as a representative forward-looking research agenda.
\end{enumerate}
Figure \ref{fig:shift} and Table \ref{tab:design-pressures} summarize the coupled shifts and the resulting design pressures.

\section{Related Work}

Our work sits at the intersection of AI for research, autonomous AI scientist systems, domain-specific scientific discovery systems, and cybersecurity agents and benchmarks. The closest prior literature contributes important building blocks, but does not yet treat cybersecurity as a scientist-level research object.

\subsection{AI for research and AI scientist surveys}
Recent surveys organize the broader landscape of AI for scientific research and AI scientist systems, covering hypothesis discovery, experiment planning, scientific writing, peer review, and the end-to-end research pipeline \citep{llm4sr_2501_04306,ai4research_2507_01903,survey_ai_scientists_2510_23045}. These works clarify how AI is reshaping research as a whole, but they remain domain-agnostic and do not engage with cybersecurity's adversarial and dual-use constraints.

\subsection{General autonomous scientist systems}
The AI Scientist demonstrated a closed-loop pipeline from ideation to paper drafting and simulated review, primarily in machine-learning subfields. The AI Scientist-v2 advances autonomy through agentic tree search and lower template dependence \citep{ai_scientist_2408_06292,ai_scientist_v2_2504_08066}. These systems show that end-to-end AI-driven research is feasible, but they are calibrated for stable computational settings rather than for adversarial research substrates.

\subsection{Domain-specific scientist systems}
Co-Scientist frames scientific collaboration as a multi-agent hypothesis-generation and proposal-development system for biomedical discovery, and Robin extends the trajectory toward hypothesis generation, experimentation, and analysis in biology \citep{ai_co_scientist_nature_2026,robin_nature_2026}. PaperQA2 shows strong scientific literature-agent performance on realistic synthesis tasks, and ERA targets expert-level empirical software generation for computational discovery \citep{paperqa2_2409_13740,era_2509_06503}. These works validate domain scientist systems, but their scientific objects are largely non-adversarial.

\subsection{Cybersecurity agents, benchmarks, and evaluation}
A separate line has rapidly advanced cybersecurity task automation. PentestGPT and HackSynth opened autonomous penetration testing, and follow-up work showed that LLM agents can exploit real one-day vulnerabilities and that agent teams can improve zero-day exploitation \citep{pentestgpt_2308_06782,hacksynth_2412_01778,one_day_vuln_2404_08144,zero_day_team_2406_01637}. More recent systems push orchestration, planning, tooling, and red-team coordination further \citep{co_redteam_2602_02164,realworld_pt_agent_2602_17622,harness_vuln_discovery_2604_20801,multiagent_pt_web_2508_20816}, while CVE-Bench, CyberGym, and multi-step attack benchmarks extend evaluation toward realistic scenarios at scale \citep{cve_bench_2503_17332,cybergym_2506_02548,multistep_attack_scenarios_2603_11214,realworld_pt_humans_2512_09882}. In parallel, evaluation and safety research has consolidated: CyberSecEval 2/3 and related work broaden cybersecurity evaluation, the Digital Cybersecurity Expert and the Cyber Defense Benchmark study role-aligned capability and open-ended defense, OpenSec targets calibration under adversarial evidence, ExploitGym benchmarks weaponized exploits at scale, and dedicated work studies offensive-security benchmarking practices and the security of agent architectures themselves \citep{cyberseceval2_2404_13161,cyberseceval3_2408_01605,digital_cyber_expert_2504_11783,cyber_defense_benchmark_2604_19533,opensec_2601_21083,exploitgym_2605_11086,benchmarking_offsec_2504_10112,security_of_ai_agents_2406_08689}.

\subsection{Frontier cyber-capable models and operational systems}
Three more recent developments mark a qualitative shift and deserve separate treatment. First, frontier models have crossed a capability threshold: Anthropic disclosed Claude Mythos Preview under Project Glasswing, a frontier model whose offensive cyber capability was strong enough to warrant restricted release through a vetted partner program, and subsequent reporting attributes large-scale defect discovery in widely deployed software to that model \citep{claude_mythos_glasswing_2026,glasswing_expansion_2026}. Second, vertical foundation models for security have begun to appear, with the Llama-3.1-based Foundation Security LLM offering an explicit cybersecurity-specialized base \citep{foundation_security_llm_2601_21051}. Third, operational AI-assisted security has moved from prototype to deployment: Microsoft's Guided Response architecture for Security Copilot documents an in-production AI system built around incident triage, action recommendation, and similar-incident retrieval over a large real-world incident corpus \citep{copilot_guided_response_2407_09017}. These are not isolated benchmarks; they are evidence that frontier cyber-capable models, vertical foundation models, and operational AI security systems are all crossing into program-scale reality at the same time, which is precisely the regime a Cybersecurity AI Scientist must be designed for.

\subsection{Gap}
Taken together, this body of work shows that cyber-agent research is already rich and methodologically dense, while AI scientist research is already mature for stable domains. What is still missing is the joint object. Existing cyber work centers on task execution, benchmarking, or agent safety; existing AI scientist work centers on stable, non-adversarial objects. Neither line treats cybersecurity as an organized scientific process in its own right. We position the Cybersecurity AI Scientist exactly in this gap.

\begin{table}[t]
\centering
\footnotesize
\begin{tabularx}{\linewidth}{@{}p{2.4cm}Y Y Y Y Y@{}}
\toprule
Line & Typical research object & Main output & Environment assumption & Governance pressure & What it still misses here \\
\midrule
AI Scientist & General scientific workflow & Ideas, code, experiments, papers & Relatively stable computational setting & Usually external or weakly embedded & Adversarial cyber object, dual-use-native workflow, digital-twin validation \\
Domain Scientist & Biology, biomed, computational discovery & Hypotheses, plans, analyses & Domain-rich but mostly non-adversarial setting & Safety matters, but usually not cyber dual-use & AI-versus-AI co-evolution, mutable cyber substrate, bounded cyber permissions \\
Cyber agents and benchmarks & Exploitation, pentest, threat hunting, cyber-role capability & Task success, benchmark scores, trajectories & Operational or benchmark environment & High, but usually task-level & Question-to-paper research system, scientist-level role-capability-artifact stack \\
Cybersecurity AI Scientist & Cybersecurity as an adversarial scientific object & Research designs, tools, traces, evaluations, governance analyses, papers & Digital twin, cyber range, and evidence pipeline as part of method & Native and internal to workflow & The target object of this paper \\
\bottomrule
\end{tabularx}
\caption{Adjacent research lines that the Cybersecurity AI Scientist distinguishes itself from.}
\label{tab:adjacent-lines}
\end{table}

\section{The Coupled Shifts: Problem and Paradigm}

Classical cybersecurity research was largely organized around human attackers, human defenders, and human analysts, with relatively stable software tools and human-paced adaptation loops. A substantial portion of the field was therefore built around human bottlenecks: expert scarcity, slow tool development, manual experiment design, and delayed evidence synthesis. AI-native conditions disturb this picture from two directions at once.

\subsection{Problem shift}
As autonomous systems increasingly participate in offensive and defensive processes, an important subset of cybersecurity problems becomes AI-versus-AI rather than human-versus-human. Some questions about analyst behavior or workflow efficiency collapse into engineering questions about orchestration and policy tuning. At the same time, new scientific questions emerge: how defensive heuristics accumulate in agent societies, how adversarial adaptation behaves under machine-speed iteration, how mutable model ecosystems reshape risk, and what counts as reliable evidence when both sides are partially autonomous. The natural unit of study moves with these questions. Security events, interaction traces, and dynamic scenarios become a more honest research handle than fixed assets and one-shot tasks. Problem definitions increasingly need to capture adaptation, coordination, containment, and evidence generation, not only task success \citep{ai4research_2507_01903,survey_agentic_cyber_2601_05293,cyber_defense_benchmark_2604_19533}.

\subsection{Paradigm shift}
The same conditions change how research is organized. Cybersecurity research moves from a human-led mode, through human-AI collaboration, toward AI-led execution under human agenda setting, where small numbers of senior researchers define questions, constraints, and acceptance criteria while large portions of execution are delegated to coordinated research agents \citep{ai4research_2507_01903,survey_ai_scientists_2510_23045}. The transition is sharper in cybersecurity than in most other domains because the substrate itself is unstable: model access policies change, tool affordances shift, guardrails evolve, evaluation environments drift, and the threat landscape mutates in response to newly available capabilities. A method that works today may become unusable tomorrow; a workflow that currently spans multiple stages may collapse into one. The substrate is non-stationary rather than steady-state, and the pipeline must be re-calibrated continuously.

The paradigm shift is also rarely a single-model story. One model may be stronger at exploitation planning, another at code synthesis, another at long-context evidence handling, and another at operating safely under stronger policy constraints. Existing evidence already shows that different models exhibit different role-aligned knowledge gaps and different end-to-end penetration-testing behavior \citep{digital_cyber_expert_2504_11783,realworld_pt_agent_2602_17622}. A credible Cybersecurity AI Scientist therefore cannot be conceived as a monolithic model endpoint. It is better understood as a calibrated orchestration layer over multiple models, tools, and evaluators whose complementary strengths must be measured and routed rather than assumed away. Governance moves with this same logic: in many scientific domains governance appears mainly as an external review layer, but in AI-native security research it becomes an internal design constraint, where permission scopes, dual-use containment, experiment isolation, and release boundaries belong inside the research method rather than around it.

\section{Defining the Cybersecurity AI Scientist}\label{sec:definition}

We define the \emph{Cybersecurity AI Scientist} as the cybersecurity-specific AI-native research object and framework that emerges once scientific-problem shift and research-paradigm shift are taken seriously together. It names the kind of research capability that becomes necessary when the object of study is adaptive, high-stakes, and tightly coupled to rapidly evolving technical infrastructure. The definition is deliberately separated from any particular organizational realization. One plausible and tractable realization is a modular multi-agent research system, in which reusable capabilities appear as bounded operators or procedural packages. Other realizations are possible; the concept does not depend on a single implementation idiom.

This separation matters. The scientific object answers \emph{what} must be studied and \emph{what kind of research system must exist}. The organizational realization answers \emph{how} such a system may be instantiated. The goal is therefore not to build a single omnipotent agent that performs every research task, and not to merely create a more capable offensive or defensive operator. It is to organize cybersecurity research as a coordinated system of specialized roles, each operating over reusable capability packages, explicit tools, bounded permissions, and structured outputs.

Under this definition, the Cybersecurity AI Scientist differs from generic AI scientist systems in at least four ways. First, it is designed for adversarial and dual-use conditions rather than stable experimental settings. Second, it must operate over a non-stationary substrate, interfacing with cyber ranges, digital twins, mutable platforms, and evidence-sensitive evaluation. Third, it should coordinate heterogeneous models and evaluators rather than presume that a single model can stably cover the workflow. Fourth, it must support not only empirical execution but also governance-aware reasoning, threat framing, and strategic hypothesis development. It is also distinct from pure attack or defense automation: a penetration-testing agent, a vulnerability-exploitation system, or a threat-hunting assistant may be useful components inside a larger framework, but none of them by itself constitutes a Cybersecurity AI Scientist. The defining feature is not task specialization but research-system organization---the ability to move from question definition to experimental design, tool assembly, controlled execution, evaluation, interpretation, and scientific reporting as a coherent process.

\subsection{Objectives: a four-zeros frame}
Concept formation is not enough; a Cybersecurity AI Scientist also needs a clear sense of what it is for. We anchor its objectives in a four-zeros frame that spans the principal dimensions along which AI-native security risk accumulates. Each zero names a distinct kind of failure that this research system is expected to study, reduce, and ultimately bound. Together they keep the framework from collapsing into either pure technology language or pure governance language. Table~\ref{tab:four-zeros} summarizes the frame.

\begin{table}[t]
\centering
\footnotesize
\begin{tabularx}{\linewidth}{@{}p{1.9cm}p{2.2cm}p{3.0cm}p{1.8cm}Y@{}}
\toprule
Dimension & Question type & Target failure mode & Primary focus & Implication for the research system \\
\midrule
Risk & Scientific & Zero hidden defects (in systems) & Machine / system & Event-driven inquiry into how flaws accumulate, propagate, and persist under AI-mediated change \\
Trust & Technical & Zero implicit trust (against human error) & Human & Calibration, role-aligned evaluation, and assistance designs that do not silently shift control \\
Incident & Operational & Zero incidents (from operational fault) & Event & Scenario-based research, replayable environments, evidence pipelines, and live-fire validation \\
Energy & Ecological & Zero loss (in organizational outcomes) & Enterprise / organization & Long-horizon agenda continuity, strategic framing, and institutional governance research \\
\bottomrule
\end{tabularx}
\caption{The four-zeros frame for the Cybersecurity AI Scientist. The four dimensions span scientific, technical, operational, and ecological failure modes, and they jointly fix the research targets that the framework must serve.}
\label{tab:four-zeros}
\end{table}

\begin{table}[t]
\centering
\footnotesize
\begin{tabularx}{\linewidth}{@{}p{3.0cm}Y Y@{}}
\toprule
Design pressure & Why a generic AI scientist pipeline is insufficient & Required response in Cybersecurity AI Scientist \\
\midrule
Adaptive adversaries & The object of study reacts to new capabilities and evaluation setups & Threat-aware framing, iterative red-blue evaluation, and bounded deployment \\
Security events as units of study & Static tasks underdescribe how cyber research is actually triggered and validated & Event-driven problem formulation and trace-centered artifacts \\
Non-stationary substrate & Models, guardrails, tools, and workflows can change during the research loop & Modular orchestration, rapid reconfiguration, and recalibration \\
Multi-model capability asymmetry & No single model is reliable across planning, coding, evidence handling, and safety-constrained operation & Model routing, evaluator diversity, and role-specialized coordination \\
Heterogeneous defense targets & Critical infrastructure, enterprise systems, platforms, and services impose different defensive constraints & Target-aware scenario design and domain-specific evaluation pathways \\
Dual-use pressure & Useful research capabilities can be repurposed offensively & Permission scopes, containment, auditability, and release boundaries \\
Evidence-intensive evaluation & Credible cyber claims depend on replayability, observability, and controlled environments & Digital twins, cyber ranges, evidence capture, and governance-linked validation \\
\bottomrule
\end{tabularx}
\caption{Design pressures that separate the Cybersecurity AI Scientist from a straightforward domain transfer of generic AI scientist pipelines.}
\label{tab:design-pressures}
\end{table}

\section{Framework Architecture}\label{sec:framework}

A systems realization of the framework can be read both as a layered architecture and as a research pipeline. The pipeline begins with agenda setting---strategic questions, scenario definitions, and governance constraints---and then moves through problem framing, literature and threat analysis, toolsmithing, controlled execution, evaluation, and scientific reporting. Across this flow sits a role layer of specialized agents: problem-framing agents, literature agents, threat-modeling agents, toolsmithing agents, experiment managers, evaluators, and reporting agents. These agents do not operate as free-form generalists. They invoke bounded procedures that encode reusable capabilities and stable interaction patterns.

Beneath the role layer sits a capability-packaging layer. Reusable capabilities are packaged as bounded operators or procedural modules with explicit input--output contracts, operational assumptions, permission scopes, and reusable routines. Three properties follow. Coordination becomes more modular and auditable, because behavior is exposed as named, contracted units rather than buried in agent policies. Partial reuse across projects becomes natural, because capabilities can be lifted without collapsing all behavior into one monolithic agent. And capability packages give governance a clean interface, because permissions can be scoped at the package level and exposure can be controlled selectively.

The next layer is the tool and runtime layer, where concrete execution happens: code generation, experiment orchestration, model routing, data handling, judge and evaluator pipelines, and runtime control logic. For cybersecurity, this layer must be coupled with an environment layer of digital twins, cyber ranges, sandboxes, replay systems, and evidence-capture pipelines. The environment is not a passive container. It is part of the research method, because the validity of cyber findings depends heavily on containment, observability, replayability, and scenario realism. Weak environments produce misleading conclusions even when nominal benchmark outcomes look strong.

The framework terminates in an artifact layer whose outputs include not only task completions but also research questions, threat models, tools, configurations, experiment traces, evaluation summaries, governance analyses, and paper-ready narrative. The Cybersecurity AI Scientist is not defined by a single score or benchmark outcome; its value lies in structured research production across the full lifecycle of inquiry. Read as design implications, four capability classes recur across these layers and any serious realization will need all of them: agenda continuity, calibration and control, high-integrity environments, and rapid tool formation.

\begin{figure}[t]
\centering
\includegraphics[width=\linewidth]{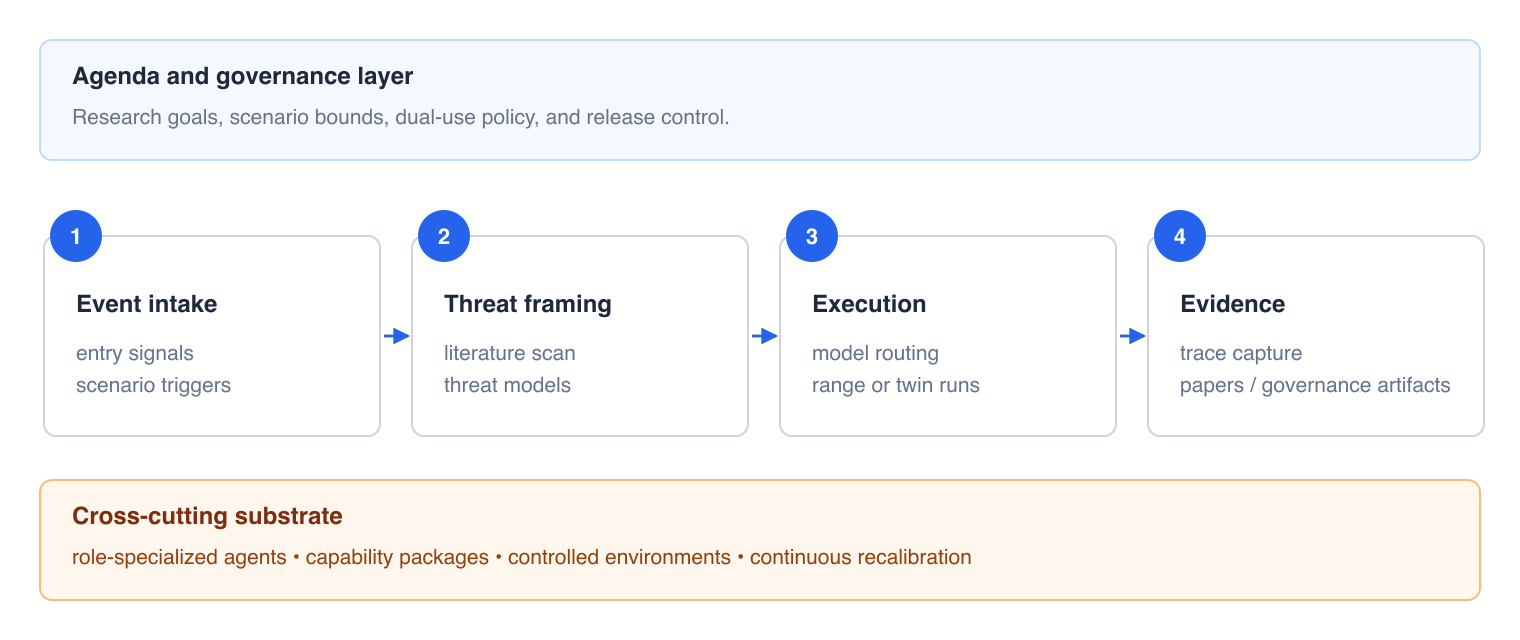}
\caption{One operational view of the Cybersecurity AI Scientist. The framework is not a single agent loop, but a governance-constrained research pipeline with role specialization, model routing, controlled environments, and evidence-producing outputs.}
\label{fig:architecture}
\end{figure}

\section{Representative Agendas under the Four-Zeros Frame}

The four-zeros frame introduced in Section~\ref{sec:definition} is not a decorative summary; it is the spine on which concrete research agendas hang. Each axis names a class of failure the Cybersecurity AI Scientist must study, organize, and ultimately bound. Each axis also corresponds to a forward-looking research agenda that is already in motion in the wider community and is sharper as a Cybersecurity AI Scientist question than as a single benchmark or single-agent problem. We walk the four axes in turn and then return to what binds them.

\subsection{Risk axis: Hidden-defect discovery under AI-mediated change}

Recent months have made it clear that frontier and vertical-cyber models can find real defects at a scale that the earlier benchmark literature was not designed to capture. Anthropic's Claude Mythos Preview, released under Project Glasswing, was withheld from general access specifically because its offensive capability was strong enough to require a vetted partner program; reported deployment has attributed large-scale defect discovery in widely deployed software, including very long-lived flaws, to that model \citep{claude_mythos_glasswing_2026,glasswing_expansion_2026}. Google's Big Sleep and DARPA's AIxCC make the same point at program scale \citep{google_summer_security_2025,aixcc_results_2025}, and the emergence of vertical foundation models for security shows that cyber-specialized substrates are now first-class \citep{foundation_security_llm_2601_21051}. Benchmarks have moved with this capability shift: CyberGym evaluates agents over more than a thousand real-world vulnerabilities in nearly two hundred open-source projects, with frontier models already reporting tens of percent single-trial success and standalone discovery of fresh zero-days, while multi-agent web-pentest systems push end-to-end exploitation \citep{cybergym_2506_02548,multiagent_pt_web_2508_20816,cve_bench_2503_17332,exploitgym_2605_11086}.

For a Cybersecurity AI Scientist, the research question on this axis is not ``can model $M$ find defect $X$?'' It is how to organize an event-driven research loop in which model capability, scenario construction, evidence handling, and governance are coordinated across many defect classes and a long research horizon. Hidden-defect discovery becomes an ongoing scientific process rather than a sequence of one-shot capability demonstrations.

\subsection{Trust axis: Calibrated assistance, multi-model routing, and role-aligned evaluation}

The Trust axis concerns the conditions under which human operators can rely on a partly autonomous research and defense system without silently surrendering control. The most explicit operational evidence here is Microsoft's Guided Response architecture for Security Copilot, which documents a production AI system organized around incident summarization, action recommendation, similar-incident retrieval, and grading---and which makes calibration against analyst behavior a first-class concern \citep{copilot_guided_response_2407_09017}. Research-side work on role-aligned cyber capability shows that different models exhibit different blind spots and different end-to-end pentest behaviors \citep{digital_cyber_expert_2504_11783,realworld_pt_agent_2602_17622}, which is exactly why the framework treats multi-model orchestration as native rather than optional. The Cybersecurity AI Scientist question here is when AI assistance starts to shift control away from operators in ways that the operators cannot detect, and how the research system itself should be designed to surface that shift rather than absorb it.

\subsection{Incident axis: Resilient agent legions and the deconstruction of terminal security}

The Incident axis is where the previous version of this paper's flagship example lives, and we keep it as a worked agenda because it makes the framework concrete. Traditional defense models assume relatively stable perimeters, department-shaped responsibility, and human-paced repair loops, and many enterprise defenses are still optimized for steady-state business environments. Those assumptions weaken once both offense and defense are mediated by autonomous or semi-autonomous agents, once attack surfaces shift faster than patch cycles, and once the very category of an ``application'' begins to dissolve into transient agent-generated behavior.

In this regime, the natural research direction is not to harden a small number of centralized defensive systems with fixed roles. It is to study what we will call \emph{resilient agent legions}: high-redundancy populations of defensive agents distributed across network boundaries, observability layers, coordination channels, and recovery functions. Such legions communicate peer-to-peer rather than only upward, exchange defensive heuristics rather than only logs, and accumulate bounded defensive experience over time. Each agent carries an \emph{event-and-defense capsule}---a packaged unit that binds a class of security events to the defensive routines that should respond to them, and that the agent itself is allowed to update as it learns. Diversity across the population, organized along a layered or fractal structure rather than a single template, is part of the design: it raises the cost of any single adaptive attack and lets the population specialize where local conditions actually differ.

\begin{figure}[H]
\centering
\includegraphics[width=\linewidth]{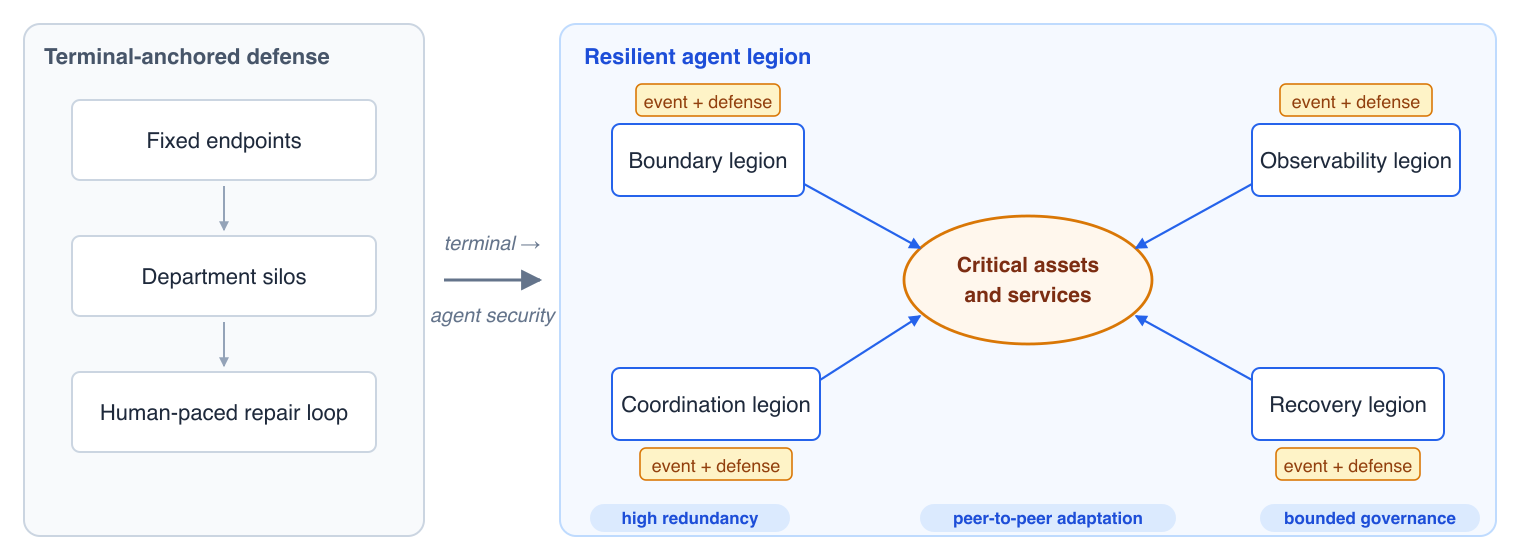}
\caption{From terminal-anchored defense to resilient agent legions. Each legion carries an event-and-defense capsule and updates it through bounded peer interaction; the protected object is no longer a fixed terminal but a governed population of defensive agents.}
\label{fig:defense-legion}
\end{figure}

This reframing carries a sharper claim. Classical \emph{terminal security} assumed a relatively stable endpoint host that could be protected by a fixed defensive product. As applications themselves become ephemeral, agent-generated, and behavior-defined, the terminal stops being the load-bearing unit. What used to be terminal security is being deconstructed into \emph{agent security}: the question is no longer how to protect a fixed endpoint, but how to organize, equip, and govern a bounded population of defensive agents whose collective behavior produces the protective effect. For mass-consumer settings, this argues for population-level monitoring rather than per-instance prevention, because the protected object will not sit still. For critical infrastructure, this argues for engineered legions whose composition, redundancy, and update behavior are themselves objects of study. The Cybersecurity AI Scientist question on this axis is no longer ``can an agent detect $X$?'' but ``how should a bounded population of defensive agents be designed, organized, evaluated, and governed under adaptive AI-mediated attack?'' Figure~\ref{fig:defense-legion} sketches the shift.

\subsection{Energy axis: Strategy, ethics, and thought experiments}

The Energy axis governs organizational-level outcomes across long horizons. A Cybersecurity AI Scientist is not only a faster experimental engine; the same shift that makes AI-native research possible also raises a class of questions that cannot be settled inside any individual experiment: how research agendas should persist as model platforms turn over, how dual-use containment should be designed when the offensive and defensive uses of a capability are difficult to separate at the level of code, and how a partly autonomous research system should be held accountable for what it chooses to study. These are strategic and ethical questions in a strict sense, not engineering details.

The framework treats such questions as native to the research method rather than as a reviewing layer placed around it. In practice this means three things. First, agenda continuity is itself an object of design: the research system should be able to carry questions, constraints, and acceptance criteria across substrate changes, so that conclusions are not silently reset every time the platform shifts. Second, ethics and dual-use review should be expressible as constraints the system can reason with at planning time, not only as post-hoc audits of finished work. Third, a portion of the research surface should be reserved for thought experiments and counterfactual studies that are difficult or unsafe to run as live experiments---questions about long-horizon misuse, about the boundary between research and operations, and about institutional placement of partly autonomous research. These complement, rather than replace, empirical work, and they belong inside the framework precisely because they shape what the empirical work is allowed to ask. This is also where the Cybersecurity AI Scientist diverges most clearly from a faster lab notebook: a lab notebook is judged by what it records; a research system of the kind we describe is judged at least as much by what it declines to ask, and by how it justifies that choice.

\subsection{From four agendas to one research object}

The four axes are not parallel case studies. Risk gives the system its sharpest empirical target---defect discovery against an adaptive adversary at frontier capability. Trust forces it to remain calibrated against the human operators and downstream decisions it touches. Incident anchors it in operational reality, where the protected object itself is shifting from terminals to bounded populations of agents. Energy holds it accountable for long-horizon organizational and ethical consequences that no single experiment can settle. A system that pursues any one of these axes well but ignores the others is a strong cyber agent, a strong assistant, a strong defense platform, or a strong governance tool---but it is not yet a \emph{Cybersecurity AI Scientist}. The defining move is to hold all four together inside one research-system organization, with the four-zeros frame as the spine and the multi-agent architecture of Section~\ref{sec:framework} as one tractable realization.

\section{Open Challenges and Boundaries}

Several challenges remain open. Defense targets are heterogeneous: what works for one class of organizational system, critical infrastructure asset, or digital platform may fail for another, and broad-transfer claims will require much stronger empirical programs than a framework paper can supply. The dual-use problem is intrinsic rather than peripheral: many capabilities required for high-quality security research are also offensively repurposable, which is why containment, release control, auditability, and permission design are central rather than ornamental. The experimental substrate is unusually demanding: cybersecurity research often needs digital twins, replayable environments, sandboxing, and high-integrity evidence capture, and poorly designed environments can produce misleading conclusions even when nominal benchmark outcomes look strong.

A further open question is institutional placement. A Cybersecurity AI Scientist may ultimately need to serve not only technical research execution but also strategic planning, governance exploration, educational cultivation, and thought-experiment-driven safety design. Whether these functions should be integrated into one system or separated into interoperable subsystems is not resolved here. We treat this as a productive open question rather than a deficiency: a framework paper that closed it prematurely would foreclose realizations we cannot yet evaluate.

\section{Conclusion}

AI-native conditions are changing not only cyber operations but cybersecurity research itself. Existing AI scientist systems make end-to-end research automation plausible. Cybersecurity agents, frontier cyber-capable models, vertical foundation models, and operational AI-security systems show that security tasks now impose distinctive pressure on capability, control, benchmarking, and governance at the same time. At their intersection, simple domain transfer is no longer a sufficient framing. Cybersecurity should be treated as a distinct scientific object for AI-driven research, one shaped by adaptive adversaries, non-stationary infrastructures, heterogeneous defense targets, and evidence-intensive validation. We define the Cybersecurity AI Scientist on that basis, anchor its objectives in a four-zeros frame spanning Risk, Trust, Incident, and Energy, and separate the object from any single organizational realization. One practical realization is a modular multi-agent research system built around bounded capability packages, explicit tools, controlled environments, and continuous calibration. Across the four axes, representative agendas---from hidden-defect discovery at frontier capability, through calibrated assistance and multi-model routing, to resilient agent legions and the deconstruction of terminal security into agent security, and on to strategic and ethical thought experiments---show why no single one of them is sufficient. The contribution of this paper is to fix the category, the design pressures, and an initial architecture on which later systems, benchmarks, and empirical programs can be built.

\bibliographystyle{plainnat}
\bibliography{40-arXiv-manuscript-v11}

\end{document}